
\centerline{Critical properties of the $XXZ$ chain in}
\medskip
\centerline{external staggered magnetic field}
\medskip

\bigskip
\medskip
\centerline{Kiyomi Okamoto${}^\dagger$ and Kiyohide
Nomura${}^{\dagger\dagger}$}
\medskip
\medskip
\centerline{Department of Physics, Tokyo Institute of Technology,}
\centerline{Oh-Okayama, Meguro, Tokyo 152, Japan}
\medskip
\bigskip
\rm
\centerline{(Received ~~~~~~~~~~September 1995)}
\bigskip
\bigskip
\bigskip
\bigskip
{\bf Abstract.}~~~We comment on the recent work of Alcaraz and
Malvezzi [1995 {\it J. Phys. A: Math. Gen.} {\bf 19} 1521] for the
critical properties of the $S=1/2$ $XXZ$ chain in staggered magnetic
field.
The method of determining the phase boundary from the finite-size
numerical data is also discussed.

\bigskip
\bigskip


\bigskip
\bigskip
\hrule
\bigskip
\bigskip
$\dagger$ E-mail address: kokamoto@stat.phys.titech.ac.jp

$\dagger\dagger$ E-mail address: knomura@stat.phys.titech.ac.jp

\vfill\eject


Recently Alcaraz and Malvezzi (AM) [1] studied the ground-state phase
diagram of the $S=1/2$ $XXZ$ spin chain in external homogeneous and
staggered magnetic fields described by
$$
    \hskip2truecm
    H(\Delta,h,h_{\rm s})
    = -{1 \over 2} \sum_{i=1}^M
      \{\sigma_i^x \sigma_{i+1}^x + \sigma_i^y \sigma_{i+1}^y
       + \Delta \sigma_i^z \sigma_{i+1}^z
       + 2[h + (-1)^ih_{\rm s}]\sigma_i^z \}
    \eqno(1)
$$
where $\sigma_i^x,\sigma_i^y$ and $\sigma_i^z$ are Pauli matrices,
$\Delta$ is the anisotropy parameter, and $h$ ($h_{\rm s}$) is the
uniform (staggered) magnetic field.
They found that the ground-state phase diagram  is composed of the
antiferromagnetic (AF) phase, the massless (ML) phase and the
ferromagnetic (FE) phase.
Although we agree their schematic phase diagram of $H(\Delta,h,h_{\rm
s})$ (figure 5 of [1]), we want to make comments on the nature of the
ground-state phase transition and also on the method to determine the
phase boundary from the finite-size numerical data.

{}~~~~~First we discuss the nature of the ground-state phase transition.
When $h \ne 0$, the uniform magnetic field breaks the spin reversal
symmetry held in the $h=0$ case, so that the nature of the phase
transition may be different from that in the $h=0$ case.
Throughout this comment we restrict ourselves to the $h=0$ case on
which AM focused.
Figure 1 shows the schematic phase diagram of $H(\Delta,h=0,h_{\rm
s})$, which is essentially the same as AM's figure 3.
They stated that the phase transition between the AF phase and the ML
phase (path~1) is of the second order.
We believe, however, it is of the infinite order, i.e., of the
Kosterlitz-Thouless (KT) type.
The operator coupled to the staggered magnetic field is irrelevant in
the ML region and is relevant in the AF region.
The excitation spectrum is either massless or massive depending on
whether this operator is irrelevant or relevant.
This mass-generating mechanism is the same as that of the sine-Gordon
model, as AM themselves denoted.
Thus the AF-ML transition of the path~1 is of the KT type, which is
seen from the well-known properties of the sine-Gordon model~[2].

{}~~~~~We can also observe the AF-ML transition along the path~2.
This transition is different from that of path~1, because it is due to
the vanishing of the coefficient (which is proportional to the
magnitude of the staggered field) of the relevant operator coupled to
the staggered magnetic field.
Thus this transition is of the second order and its critical exponents
vary continuously.

{}~~~~~Next we discuss on the method to determine the AF-ML phase
boundary from the finite-size numerical data obtained by the numerical
diagonalization of the Hamiltonian.
AM used the $M \rightarrow \infty$ extrapolation of the sequences
$(\Delta^{(M)}, h^{(M)},h_{\rm s}^{(M)})$ $(M = 2,4,\cdots)$ obtained
by solving the so-called phenomenological renormalization group (PRG)
equation$$
    \hskip2truecm
    MG_M(\Delta,h,h_{\rm s}) = (M-2)G_{M-2}(\Delta,h,h_{\rm s})
    \eqno(2)
$$
where $G_M(\Delta,h,h_{\rm s})$ is the gap of the Hamiltonian (1) with
$M$ sites.
At the fixed point of the PRG equation (2), the gap $G_M$ behaves as
$$
    \hskip2truecm
    G_M \sim M^{-1}
    \eqno(3)
$$
in the lowest order of $M^{-1}$.
If the transition is of the second order, the PRG method leads to the
correct transition point because the system is massless and equation
(3) holds only at the transition point.
In case of the KT transition, on the other hand, care must be taken
for the application of the PRG method.
Since the present system is massless not only at the AF-ML transition
line but also in the whole of the ML region, the PRG relation (2) is
satisfied in the lowest order of $M^{-1}$ in the whole of the ML
region.
Where is the fixed point of the PRG equation?
It is controlled by the lowest order correction to equation (3) which
may comes from the operator coupled to the staggered magnetic field.
Thus the fixed point of the PRG equation locates at the point where
the staggered field vanishes.
If this is the case, the transition point obtained through the  PRG
method is brought over from the AF-ML point to the $h_{\rm s} = 0$
line.
Then the simple application of the PRG method to the KT transition is
dangerous.
In the present problem, of course, there may be other corrections
which make the situation more complicated.

{}~~~~~Let us demonstrate that the PRG solution may lead to an incorrect
critical point for the KT transition~[3].
When $h = h_{\rm s} = 0$, as is well-known, the Hamiltonian (1) is
exactly solvable by the use of the Bethe ansatz method.
Its ground-state is either the AF state or the ML state depending on
whether $\Delta < -1$ or $-1 \le \Delta <1$.
The excitation gap in the AF state behaves as~[4,5]
$$
    G(\Delta)
    \simeq 8\pi
      \exp\left( - {\pi^2 \over 2\sqrt{2(|\Delta|-1)}} \right)
    ~~~~~(\Delta \rightarrow -1-0)
    \eqno(4)
$$
which indicates that this AF-ML transition at $\Delta = -1$ is of the
KT type.
If we apply the PRG method to the finite-size numerical data of the
excitation gap, we obtain $\Delta_{\rm c} = -0.50706$ $(M = 10,12)$
and $\Delta_{\rm c} = -0.47564$ $(M = 18,20)$.
Therefore the critical value of $\Delta_{\rm c}$ obtained from the PRG
equation goes far off from the exact value $\Delta_{\rm c} = -1$ as
$M$ increases.
Where is the fixed point of the PRG equation in this case?
Since the mass in the AF state is generated by the operator coming
from the Umklapp scattering between the Jordan-Wigner fermions
originated from the $S_i^z S_{i+1}^z$ term in the spin Hamiltonian,
the fixed point is the $XY $ point ($\Delta = 0$) where there is no
$S_i^z S_{i+1}^z$ term resulting in the vanishment of the interaction
between fermions.
Thus the PRG solution is brought over from the true transition point
$\Delta_{\rm c} = -1$ to the $XY$ point.
This example was also noticed by Bonner and M\"uller~[6] and by
S\'olyom and Ziman~[7].

\vfill\eject


\parindent25truept
\noindent
{\bf References}

\item{[1]} Alcaraz F C and Malvezzi A L 1995 {\it J. Phys. A: Math.
Gen} {\bf 28} 1521

\item{[2]} for instance, Kogut J B 1979 {\it Rev. Mod. Phys.} {\bf 51}
659

\item{[3]} Nomura K and Okamoto K 1994 {\it J. Phys. A: Math. Gen}
{\bf 27} 5773

\item{[4]} des Cloizeaux J and Gaudin M 1966 {\it J. Math. Phys} {\bf
7} 1384

\item{[5]} Yang C N and Yang C P 1966 {\it Phys. Rev.} {\bf 151} 258

\item{[6]} Bonner J C and M\"uller G 1984 {\it Phys. Rev.} B {\bf 29}
5216

\item{[7]} S\'olyom J and Ziman T A L 1984 {\it Phys. Rev.} B {\bf 30}
3980

\vfill\eject


\parindent0pt

{\bf Figure caption}
\bigskip
Figure 1: Schematic ground-state phase diagram of the Hamiltonian
$H(\Delta,h=0,h_{\rm s})$.
The antiferromagnetic, massless and ferromagnetic phases are indicated
by AF, ML  and FE, respectively.
The AF-ML transition along the path 1 is of the KT type and that along
the path 2 is of the second order.
The transition to the FE state is of the first order.

\end